\documentclass[11pt]{iopart}

\usepackage{amsgen}
\usepackage{amssymb}
\usepackage{graphicx}
\usepackage{setstack}
\usepackage{color}
\usepackage{hyperref}

\begin{document}

\title[]{Post-Newtonian Global Conservation Laws in the Presence of Sensitive Bodies}

\author{Timothy Clifton$^{1*}$, Asta Heinesen$^{2*+}$ and Oliver Pitt$^{3*}$}
\address{${\phantom{.}}^*$Department of Physics \& Astronomy, Queen Mary University of London, UK.}
\address{${\phantom{.}}^+$Niels Bohr Institute, Blegdamsvej 17, DK-2100 Copenhagen, Denmark.}
\ead{$^1$t.clifton@qmul.ac.uk, $^2$asta.heinesen@nbi.ku.dk, $^3$o.pitt@qmul.ac.uk}

\vspace{0.75cm}
\begin{abstract}
The parameterized post-Newtonian (PPN) approach is the state of the art formalism for performing theory independent tests of weak-field gravity, and for constraining possible deviations from Einstein's theory. Within this framework, global conservation laws are useful for the calculation of dynamics and for giving meaning to parameters. In this paper we extend the concept of semi-conservative and fully-conservative theories of gravity to include situations in which compact astrophysical bodies are modeled as masses that are sensitive to their local environment, as relevant for theories that violate the strong equivalence principle. We find that globally conserved quantities can still exist in the presence of such sensitivities, and find their explicit forms when they do. We identify new ways of writing the coefficients that enter into the PPN metric when a theory of gravity admits conserved quantities in the presence of a sensitive body, and demonstrate the applicability of our approach by comparing it to known results in scalar-tensor theories of gravity.
\end{abstract}

\vspace{0.75cm}
\section{Introduction} 

The parameterized post-Newtonian (PPN) approach is the flagship formalism for interpreting the results of gravitational physics experiments and observations in weak-field systems, and for quantifying possible deviations from the predictions of Einstein's theory of general relativity. The first pioneering steps towards this formalism were made by Eddington in 1922 \cite{eddington1923mathematical}, with the detailed modern version constructed by Nordvedt and Will in and around the 1970s \cite{nordtvedt1968equivalence, will1971theoretical, will1972kenneth, will1973relativistic, clifford1981theory}. It is built on a post-Newtonian expansion, which is a slow-motion and weak-field approximation used for extracting the mildly relativistic and non-relativistic limits of metric-based theories of gravity. The parameters that appear in the PPN formalism quantify the amplitude of each possible type of gravitational potential that will be generated by matter in a given theory, and can be probed by observations and experiments in order to constrain deviations from Einstein's theory \cite{will2014confrontation}. 

Such an approach is directly applicable to all relativistic phenomena in the Solar System, but not to strong-field systems such as the interior of Neutron stars or the vicinity of black holes. In these cases one could still hope to model the distant gravitational field as being post-Newtonian, but should not attempt to apply this formalism to the strong-field region directly. One then needs an effective version of the energy-momentum tensor to act as a source for the distant post-Newtonian fields of these strong-field objects, which leads to something of a conundrum; one needs information about the strong-field behaviour of a theory in order to be able to model its weak-field. Of course, this problem becomes increasingly pressing with the advent of gravitational wave astronomy, in which both gravitational and electromagnetic radiation from the inspiral and merger of compact bodies is directly detectable \cite{bailes2021gravitational}.

Within Einstein's theory this problem is resolved by the strong equivalence principle (SEP), which states that massive self-gravitating objects follow the same trajectories as massless test particles. In the case of scalar-tensor theories of gravity, where the SEP is no longer valid, a solution as to how to model the weak-field of compact bodies was provided by Eardley \cite{eardley1975observable}. This was achieved by allowing the effective mass that enters into the post-Newtonian description of the distant weak field to depend on the local value of the gravitational scalar field $\varphi$ at the location of the compact body. As the mass of a body is the principle source of its gravitational field in the post-Newtonian limit, and as it is possible to locally transform away all components of the gravitational field except for tidal force tensors and the value of $\varphi$, it was argued that this approach was sufficient to model the main contribution to the distant gravitational field of compact bodies within this class of theories. This approach provides a simple fix to a complicated problem, and is described in the literature on this subject as introducing a ``sensitivity'' into the mass of the compact object.

In a recent paper, two of us extended this programme so that it could be understood in a theory-independent context (i.e. not just within the scalar-tensor class of gravitational theories) \cite{pitt2026constraining}. This was achieved by allowing the mass of compact objects to depend not on the invariant value of a fundamental gravitational scalar field (which may not exist in general), but on {\it all} of the scalars that exist in the pantheon of PPN gravitational potentials. Assuming that the fields to which the mass of a compact body is sensitive can be expressed in terms of these potentials, and working in standard post-Newtonian gauge, this approach allowed us to generalize the concept of compact body sensitivities to a much wider class of theories, which thereby allowed us to calculate the two-body dynamics of compact objects without specifying a theory of gravity.

This approach to modeling the weak-field of compact objects fits well with the general philosophy of the standard PPN formalism; it allows gravitational physics calculations to be performed in a theory-independent way, so that observations and experiments can be used to constrain a pre-specified set of parameters that have direct physical interest for the problem at hand. It can be thought of as an extension of the standard PPN approach so that it can be applied to neutron stars and black holes, which is of particular interest given the rapidly developing field of gravitational wave astronomy \cite{bailes2021gravitational}. It can also be used to constrain the cosmological time-evolution of PPN parameters \cite{pitt2026constraining, Sanghai:2016tbi, Clifton:2018cef, Anton:2021vsr, Thomas:2022phf, Thomas:2024bsv, Anton:2025zer, Thomas:2025qiz}, thus providing a theory independent route for constraining deviations from general relativity over cosmological scales. This extended formalism brings with it the requirement to include a large number of new sensitivity parameters, as we will discuss below. 

Our goal in this paper is to consider {\it global conservation laws} -- such as those involving total mass, momentum and angular momentum -- in the presence of our new set of theory-independent sensitivity parameters. Such laws have a long history in the theory of relativistic gravitational physics \cite{chandrasekhar1965post}, and are of use for classifying theories of gravity as being {\it fully-conservative}, {\it semi-conservative} or {\it non-conservative} \cite{clifford1981theory}. They are important properties for gravitational systems and theories to admit, and can be shown to be linked to the existence of a covariant formulation for the gravitational Lagrangian of metric-based theories of gravity \cite{lee1974conservation}. Here we will consider global conservation laws in the presence of the theory-independent sensitivities of compact objects described above. We will derive the conditions under which globally-conserved quantities can exist in the presence of a sensitive body, and give detailed expressions for them when they do.

In Section 2 we give the relevant background theory for this work, including on the PPN approach, theory-independent sensitivities, and global conservation laws. In Sections 3 and 4 we determine conservation equations to leading and higher post-Newtonian {orders}, and in Section 5 we investigate the conditions for the existence of globally-conserved integral quantities in the presence of theory-independent sensitivities. We then discuss an example theory and some special cases in Section 6, before concluding in Section 7. We work in units such that $c=G=1$, and use Greek indices $\{\mu, \nu, \sigma, \dots\}$ for space-time components, with $0$ reserved for the time component and Latin indices $\{i, j, k, \dots\}$ reserved for space. {We use overdots $\dot{S}$ to denote time-derivatives of a quantity $S$, and $S_{,i} = \partial_i S$ to denote spatial derivatives, with  $\nabla^2 S = \partial^i \partial_i S$ denoting the spatial Laplacian operator.} We choose to work {with a counting scheme} such that $\rho \sim v^2$ in our post-Newtonian expansions, and so that time derivatives add an extra order-of-smallness in $v$ to any quantity on which they act.

\section{Background Theory}

In this section we will introduce the relevant background theory for the study of both the PPN approach and the inclusion of sensitivities in the treatment of compact bodies.

\subsection{Parameterized post-Newtonian approach}

The PPN formalism is based on a post-Newtonian expansion of Minkowski space, in which the perturbation to the metric is assumed to be small,
\begin{equation}
g_{\mu \nu} = \eta_{\mu \nu} + h_{\mu \nu} \qquad {\rm where} \qquad \vert h_{\mu \nu} \vert \ll 1 \, ,
\end{equation}
and where 3-velocities and time-derivatives are also taken to be small,
\begin{equation}
\vert{v^i} \vert \ll 1 \qquad {\rm and} \qquad \frac{\partial/\partial t}{\partial/ \partial {x^i}} \ll 1 \, .
\end{equation}
It is then supposed that the metric can be written in terms of the following set of gravitational potentials: 
\begin{equation} \label{pots}
\left\{U , X, V^i, \Phi_1, \Phi_2, \Phi_3, \Phi_4, \Phi_5, \Phi_6 , \Phi_{\rm W} \right\} \, ,
\end{equation}
where the first is the Newtonian gravitational potential{, $U \sim v^2$,} that satisfies $\nabla^2 U = -4 \pi \, \rho^*$, and $\rho^{\star}= \sqrt{g} \,u^0\rho$ is the (locally) conserved mass density written in terms of the time component of the matter 4-velocity $u^{\mu}$ and the mass density $\rho$. The quantity {$X \sim v^2$} is the superpotential satisfying $\nabla^2 X = 2 U$, and $V^i$ and the $\Phi_i$ are post-Newtonian potentials of orders $v^3$ and $v^4$, respectively. The detailed mathematical form of each of these potentials is given in Appendix A {and Reference \cite{clifford1981theory}, and have the following physical interpretations: $V^i$ is the vector gravitational field of the momentum density ($\rho^* v^i$), $\{\Phi_1$, $\Phi_2$, $\Phi_3$, $\Phi_4\}$ are (respectively) the gravitational fields of the kinetic energy density ($\rho^*v^2$), the gravitational potential energy density ($\rho^* U$), the internal energy density ($\rho^* \Pi$) and the pressure ($p$), while $\{\Phi_5, \Phi_6 , \Phi_W\}$ have less direct meanings.} 

In the PPN approach they are combined in the components of the metric in the following way:
\begin{eqnarray} \label{g00}
g_{00} &=& -1 +2 \, \alpha \, U + 2(\psi-\beta\, U^2) +O(v^6) \\ \label{g0i}
g_{0i} &=& -\gamma_1 V_i - \gamma_2 X_{,0i} +O(v^5) \\ \label{gij}
g_{ij} &=& (1+2 \, \gamma\, U) \, \delta_{ij} +O(v^4) \, ,
\end{eqnarray}
where
\begin{equation} \label{psi}
\psi = \beta_1 \, \Phi_1 + \beta_2 \, \Phi_2 + \beta_3 \, \Phi_3 + \beta_4 \, \Phi_4 + \beta_6 \, \Phi_6 + \beta_W \, \Phi_W \, . 
\end{equation}
The coefficients $\{\alpha, \beta, \gamma, \gamma_1, \gamma_2, \beta_1, \beta_2, \beta_3, \beta_4, \beta_5, \beta_6, \beta_W \}$ are {\it a priori} unspecified, such that the metric has been written as the most general possible polynomial combination of the stated potentials that maintains the correct index structure and dimensionality up to the required order of accuracy. Identities between potentials, such as $\ddot{X} = \Phi_1+2 \Phi_4 - \Phi_5 - \Phi_6$, have been used to remove certain possible contributions to the metric, and a choice of coordinates has been made to diagonalize $g_{ij}$ and to remove $\Phi_5$ from $g_{00}$. This is the `standard post-Newtonian' form of the PPN metric, which is explained in further detail in Reference \cite{clifford1981theory}. 

{The reader may note that Equations (\ref{g00})-(\ref{psi}) are not in their standard form, as given in e.g. Reference \cite{clifford1981theory}. In particular, we have not set $\alpha=1$ by choice of units, as it is useful to keep this parameter explicit for astrophysical systems at large cosmological redshifts (see e.g. \cite{Thomas:2025qiz}). The parameter $\{\beta_2 \}$, which is the coupling parameter of the gravitational field $\Phi_2$, will require some adjustment here, as sensitivities will be seen to affect the gravitational potential energy densities that source this potential. We deal with this explicitly in Section \ref{sec:semi}. The other \{$\beta_i$\}, and the parameters $\{\beta, \gamma, \gamma_1, \gamma_2\}$, will all eventually be given by their standard forms, which are given in Equations (\ref{b1})-(\ref{fW}) and below Equation (\ref{aeq}). For now we will keep them in the form above though.
}

\subsection{Theory-independent sensitivities}

A theory-independent notion of sensitivity can be given to a compact body by {representing its} mass density as a function of the gravitational fields at its location{, such that $\rho=\rho(U,\Phi_i, \dots)$ where $\Phi_i$ are the post-Newtonian gravitational potentials from Equation (\ref{pots}). The basic idea here is to account for the effects that strong-equivalence-principle violations should have on the gravitational interactions of a compact body by allowing its mass to be determined by its local environment. As most alternative theories of gravity are expected to violate the strong equivalence principle, so compact bodies in most theories should be expected to represented by mass densities that are sensitive in this way. This approach generalizes the formalism devised by Eardley for modeling compact bodies in scalar-tensor theories \cite{eardley1975observable}, thereby allowing it to be applied to a much wider class of metric theories of gravity \cite{pitt2026constraining}.}

{In weak field environments, the field in Equation (\ref{pots}) all have magnitudes that are much less than unity. As such, we can Taylor expand the energy density so that it is given by:} 
$$
\rho = \rho_0 +\rho_{,U} \, {U} +\frac{1}{2} \rho_{,UU} \, U^2 + \rho_{,\Phi_i}\, \Phi_i + \dots \, ,
$$
where the bare mass density $\rho_0$ is {intended} to be evaluated at a point where all gravitational potentials vanish. In general, the mass of a compact body {might} be expected to be a function of any of the metric potentials that exist at its location, and here we will let it be a function of any of the dimensionless scalar potentials that appear in the PPN approach, including the Whitehead potential $\Phi_W$ and the potential $\Phi_5$ (which is in general non-vanishing even though it can be removed from the metric component $g_{00}$ by a coordinate choice{, as in the `standard post-Newtonian gauge'\footnote{{In post-Newtonian theory one has the gauge freedom to change coordinates as $x^{\mu} \rightarrow x^{\mu} +\xi^{\mu}$, where $\xi^0 \sim v^3$ and $\xi^i \sim v^2$. This freedom can be used to diagonalize the leading-order part of $h_{ij}$ and to remove $\Phi_5$ from $h_{00}$, which results in ``standard post-Newtonian gauge'' (see Reference \cite{clifford1981theory} for details).}}.}).

It is now convenient to introduce the following sensitivity parameters:
\begin{equation}
s_{U} \equiv \frac{\partial  \ln \rho}{\partial U} \, , \qquad s'_{U} \equiv \frac{\partial^2  \ln \rho}{\partial U^2} \qquad {\rm and} \qquad s_{\Phi_i} \equiv \frac{\partial  \ln \rho}{\partial \Phi_i} \, ,
\end{equation}
where the index $i$ picks out a potential from the set $\left\{\Phi_1, \Phi_2, \Phi_3, \Phi_4, \Phi_5, \Phi_6 , \Phi_{\rm W}
\right\}$. 
If we now build a stress-energy tensor in the perfect fluid form,
\begin{equation}
T^{\mu \nu} = \rho (1+\Pi) u^{\mu} u^{\nu} +p \left(u^{\mu} u^{\nu} + g^{\mu\nu} \right) \, ,
\end{equation}
assuming that the bare mass $\rho_0$ obeys the usual local covariant energy-momentum conservation equations, as should be expected {for fluid quantities that are not dependent on sensitivity parameters}, then gives 
\begin{eqnarray} \label{dT}
\hspace{-1.5cm} T^{\mu\nu}_{\phantom{\mu\nu} ;\nu} &=& 
- \rho^{\star} g^{\mu\nu} \left[ s_{U}  \left(1-\frac{1}{2} v^2 -3 \gamma U \right)  U_{,\nu}
+\left(s_U^2 + s_U'\right) U U_{,\nu} 
+ \sum_i s_{\Phi_i} \Phi_{i,\nu} \right] \, ,
\end{eqnarray}
where $\rho^{\star} \equiv \rho \left(1+\frac{1}{2} v^2 +3 \gamma U\right)$. In the absence of sensitivities we recover $T^{\mu \nu}_{\phantom{\mu\nu} ;\nu}=0$, but for any of $\{s_U, s'_U, s_{\Phi_i}\}$ non-zero we have in general that $T^{\mu \nu}_{\phantom{\mu\nu} ;\nu} \neq 0$. This does not imply any lack of covariant conservation in the full field equations of theories of gravity with compact bodies, but does show that when describing compact bodies as sensitive (for the purposes of calculating their distant weak-field effects) we have that they must be described by an effective energy-momentum tensor that is {\it not} covariantly conserved\footnote{We note that in general relativity we expect $T^{\mu \nu}_{\phantom{\mu\nu} ;\nu}=0$ directly from the Bianchi identities, which means that sensitivities cannot be non-zero in that theory, in keeping with the strong equivalence principle. This does not follow in other theories (see e.g. Ref. \cite{Mirshekari2013compact}). {More generally, energy-momentum conservation also does not follow from diffeomorphism invariance of the matter action, as it is a function of both the metric and the environmental fields given in Eq. \ref{pots} (see e.g. Appendix B of Ref. \cite{pitt2026constraining})}.}.

Our goal in what follows will be to study globally-conserved quantities at post-Newtonian order of accuracy in the presence of the sensitivities described above. This will include the conditions required for the existence of conserved quantities of various types, as well as a derivation of their explicit form.

\subsection{Global conservation laws}

Our approach to identifying global conservation laws will closely follow that of Will \cite{clifford1981theory}, which is based on a generalization of the ideas used by Chandrasekhar to establish globally-conserved quantities at post-Newtonian order in general relativity \cite{chandrasekhar1965post}. The idea is to find a quantity $\tau^{\mu \nu}$ that is conserved, such that 
\begin{equation} \label{dt0}
\tau^{\mu \nu}_{\phantom{\mu\nu} , \nu}=0 \, .
\end{equation}
The reader may note that this equation involves partial derivatives, not covariant derivatives, as we are looking for quantities that are conserved under standard partial derivatives with respect to coordinate position and time. For isolated systems, and assuming that $\tau^{\mu\nu}$ is symmetric, the existence of an equation of the form given in (\ref{dt0}) leads to the following global conservation laws \cite{thorne2000gravitation}:
\begin{equation}
\frac{{\rm d} P^{\mu}}{{\rm d}t}=0 \qquad {\rm and} \qquad \frac{{\rm d}J^{\mu \nu}}{{\rm d}t} =0 \,,
\end{equation}
where
\begin{equation} \label{pandj}
P^{\mu} = \int \tau^{\mu 0} {\rm d}^3x \qquad {\rm and} \qquad J^{\mu\nu} = 2 \int x^{[\mu} \tau^{\nu]0} \rm{d}^3x \, 
\end{equation}
are the conserved energy and linear momentum, and the conserved angular momentum and centre-of-mass momentum, respectively. If $\tau^{\mu \nu}$ exists, but is not symmetric, then only $P^{\mu}$ is conserved, with ${\rm d} J^{\mu\nu} /{\rm d}t = -2 \int \tau^{[\mu\nu]} {\rm d}^3x$ \cite{clifford1981theory}.

The quantity $\tau^{\mu \nu}$ should incorporate the physics of relativistic gravitational fields, and in the absence of gravity should reduce to the stress-energy tensor. This motivates the form
$$ \tau^{\mu \nu} \propto T^{\mu \nu} + t^{\mu \nu} \,,$$
where $t^{\mu \nu}$ is constructed from gravitational fields only, and where the constant of proportionality reduces to unity when gravity is absent. Lee, Lightman and Ni have shown that such objects exist under quite general conditions; for example, every Lagrangian-based covariant metric theory of gravity without prior geometry is known to admit a $t^{\mu \nu}$ (though not necessarily a symmetric one), in which the conserved quantities are associated with invariance under coordinate transformations \cite{lee1974conservation}. Here we wish to determine the extent to which such quantities exist when matter is sensitive.

As $U$ is the only metric potential at order $v^2$, we can write that to first post-Newtonian order we have
\begin{equation} \label{tau}
\tau^{\mu \nu} = (1-a\, U) (T^{\mu \nu} + t^{\mu \nu}) \, ,
\end{equation}
where $a$ is some constant to be determined. Written in this form, Equation (\ref{dt0}) implies
\begin{equation} \label{dt}
t^{\mu \nu}_{\phantom{\mu \nu},\nu} - a \, U_{,\nu} t^{\mu\nu} = a\, U_{,\nu} T^{\mu \nu} + \Gamma^{\mu}_{\phantom{\mu} \nu \rho} T^{\rho \nu} + \Gamma^{\nu}_{\phantom{\nu} \nu \rho} T^{\rho \nu} - T^{\mu\nu}_{\phantom{\mu\nu} ;\nu} \, .
\end{equation}
At this stage, and in the absence of sensitivities, the conservation of stress-energy is usually used to remove the last term in the equation above. Our programme will be to look for solutions of Equation (\ref{dt}), using the expression for $T^{\mu\nu}_{\phantom{\mu\nu} ;\nu}$ given in Equation (\ref{dT}), in the presence of a sensitive compact body.

\section{Conservation equations at leading orders} \label{sec:leading}

We can see from Equation (\ref{tau}) that $\tau^{\mu \nu}$ will have contributions at
\begin{equation} \label{tausize}
\tau^{00} \sim v^2 \, , \qquad \tau^{0i} \sim \tau^{i0} \sim v^3 \qquad {\rm and} \qquad \tau^{ij} \sim v^4 \, ,
\end{equation}
as these are the sizes of contributions to the various components of the stress-energy tensor, $T^{\mu \nu}$. The addition of $t^{\mu \nu}$, which obeys Equation (\ref{dt}), will contribute leading-order terms of size 
\begin{equation} \label{tsize}
t^{00} \sim v^4 \, , \qquad t^{0i} \sim t^{i0} \sim v^5 \qquad {\rm and} \qquad t^{ij} \sim v^4 \, .
\end{equation}
This means that the leading-order contributions to $\tau^{\mu \nu}$ will be those indicated in (\ref{tausize}), with contributions from $t^{\mu \nu}$ at the orders indicated in (\ref{tsize}) needed to determine the leading-order contribution of the gravitational field to any globally conserved quantities.

In this section we will calculate the temporal and spatial components of Equation (\ref{dt}) inside a sensitive body to orders $v^4$ and $v^5$, respectively. This will be sufficient to find the components of $\tau^{00}$, $\tau^{0i}$ and $\tau^{ij}$ up to orders $v^4$, $v^5$ and $v^4$, which will correspond to the order beyond leading in these quantities. In the next section this will be built on to find $\tau^{i0}$ to order $v^5$, and $\tau^{ij}$ to order $v^6$, as well as integrability conditions for globally conserved quantities to exist at the required orders of accuracy in the presence of a sensitive compact object.

\subsection{Finding $\tau^{ij}$ to order $v^4$}

To find $\tau^{ij}$ to the required order we can consider the spatial components of Equation (\ref{dt}), which give
\begin{equation}
t^{i \mu}_{\phantom{i\mu},\mu} =  t^{i0}_{\phantom{i0},0} +t^{ij}_{\phantom{ij},j}  = -\tilde{\alpha} \, \rho^* \, U_{,i} +O(v^6) \, ,
\end{equation}
where $\tilde \alpha \equiv \alpha - s_U$. Using the definition of $U$, which is given implicitly by $\nabla^2 U =- 4 \pi \, \rho^*$, we can write this as
\begin{equation}
4 \pi \, t^{i \mu}_{\phantom{i\mu},\mu} = \tilde{\alpha} \, \partial_j \left[U_{,i}U_{,j} - \frac{1}{2} \delta_{ij} \vert \nabla U \vert^2 \right] +O(v^6) \, ,
\end{equation} 
from which we can identify that $t^{i0} \sim v^5$ (at most) and that
\begin{equation} \label{tij4}
4 \pi \, t^{ij} = \tilde{\alpha} \left(U_{,i}U_{,j} - \frac{1}{2} \delta_{ij} \vert \nabla U \vert^2 \right) + O(v^6) \, .
\end{equation}
This gives
\begin{equation}
\tau^{ij} = \rho^* v^i v^j  {+p \, \delta^{ij}} +\frac{\tilde{\alpha}}{4\pi} \left(U_{,i}U_{,j} - \frac{1}{2} \delta_{ij} \vert \nabla U \vert^2 \right) + O(v^6) \, ,
\end{equation}
which is the result we required from this section, and shows how sensitivity enters into the expression for $\tau^{ij}$ at this order of accuracy: in the replacement $\alpha \rightarrow \tilde{\alpha}$.

\subsection{Finding $\tau^{00}$ to order $v^4$ and $\tau^{0i}$ to order $v^5$}

We now need to consider the temporal component of Equation (\ref{dt}) to order $v^5$, which we find results in 
\begin{eqnarray}
\hspace{-1.5cm}
t^{0\mu}_{\phantom{\mu},\mu} &=&  t^{00}_{\phantom{00},0} + t^{0i}_{\phantom{0i},i} 
=
 -(2 \tilde {\alpha}-3 \tilde{\gamma} -  \tilde{a}) \rho^*\, U_{,0} -  (3 \tilde {\alpha}-3 \tilde{\gamma} -  \tilde{a}) \rho^* \, U_{,i} v^i +O(v^7) \,,
\end{eqnarray}
where $\tilde{\gamma} \equiv \gamma+s_U$ and $\tilde{a} \equiv a - 6 s_U$. Now, using the implicit definition $\nabla^2 V^i = - 4\pi \, \rho^* \, v^i$ and the result $U_{,0} = - V^i_{\phantom{i},i}$, which follows from the leading-order part of the local energy conservation equation and the definitions of $U$ and $V^i$, we can write this as
\begin{eqnarray}
\hspace{-1.5cm}
4 \pi \, t^{0\mu}_{\phantom{\mu},\mu} &=& \partial_i \left[ (2 \tilde{\alpha}-3 \tilde{\gamma}  - \tilde{a}) U_{,i} U_{,0} + 2 (3 \tilde{\alpha}-3 \tilde{\gamma}  - \tilde{a}) U_{,j} V_{[j ,i]} \right] 
\\ &&\hspace{4.5cm}- \left( \frac{5}{2} \tilde{\alpha}-3 \tilde{\gamma}  - \tilde{a} \right) \partial_t  \vert \nabla U \vert^2 +O(v^7) \, . \nonumber
\end{eqnarray}
We can then read off
\begin{eqnarray} \label{t00}
4\pi\, t^{00} &=&- \left(\frac{5}{2} \tilde{\alpha}- 3 \tilde{\gamma} -  \tilde{a} \right) \vert \nabla U \vert^2 + O(v^6) \\ \label{t0i}
4\pi \,t^{0i} &=& (2 \tilde{\alpha}-3 \tilde{\gamma}  - \tilde{a}) U_{,i} U_{,0} + 2 (3 \tilde{\alpha}-3 \tilde{\gamma}  - \tilde{a}) U_{,j} V_{[j ,i]} +O(v^7) \, .
\end{eqnarray}
This gives
\begin{eqnarray} \label{tau00}
&&\hspace{-2.5cm}\tau^{00} = \rho^* \left(1-\tilde{a} U+\Pi {+\frac{1}{2} v^2 +(2 \tilde{\alpha} -3 \tilde{\gamma}) U }\right) -\frac{1}{4\pi}\left( \frac{5}{2} \tilde{\alpha}-3 \tilde{\gamma} -  \tilde{a} \right) \vert \nabla U \vert^2 + O(v^6) 
\\ \label{tau0i}
&&\hspace{-2.5cm}\tau^{0i} = \rho^* v^i \left( 1 {-\tilde{a}U} +\Pi +\frac{1}{2} v^2{+(2 \tilde{\alpha} -3 \tilde{\gamma}) U }\right) +p\, v^i \\ \hspace{-1.5cm}
&&\hspace{1.5cm}+ \frac{1}{4\pi} (2 \tilde{\alpha}-3 \tilde{\gamma}  - \tilde{a}) U_{,i} U_{,0} + \frac{1}{2\pi} (3 \tilde{\alpha}-3 \tilde{\gamma}  - \tilde{a}) U_{,j} V_{[j ,i]} +O(v^7) \, , \nonumber
\end{eqnarray}
which is our next result: in these quantities the effect of including sensitivity is to replace $\{\alpha, \gamma, a \} \rightarrow \{\tilde{\alpha}, \tilde{\gamma}, \tilde{a}\}$, with no further impositions required. To go further we now need to consider the next-to-leading-order parts of the spatial component of Equation (\ref{dt}).

\section{Conservation equations at higher order} \label{sec:next}

To calculate conservation equations at higher order we will need the post-Newtonian potential $\psi$, from Equation (\ref{psi}). This will need to be adapted from its standard form by taking into account the change in gravitational field due to the sensitivity of the compact body, which will require some extra consideration of the following potential: 
\begin{equation}
\Phi_2 \equiv \int \hspace{-5pt}\int \frac{\rho^{\star \prime}}{\vert {\bf x}-{\bf x'}\vert} \frac{\rho^{\star \prime \prime}}{\vert {\bf x'}-{\bf x''} \vert} {\rm d}^3 x'{\rm d}^3 x'' \, .
\end{equation}
This potential will have support from within the sensitive body, through the integral over ${\bf x^{\prime \prime}}$, and can be understood as the gravitational field of the gravitational potential energy density. This means that the potential $\Phi_2$ will depend on both sensitive and insensitive bodies, and will therefore require special treatment.

We resolve this complication by making the following substitution into the standard expression for $\psi$, as given in Equation (\ref{psi}): 
\begin{equation} \label{replace}
\beta_2 \Phi_2 \rightarrow \beta_2^{({\rm S})} \Phi_2^{({\rm S})} + \beta_2^{({\rm I})} \Phi_2^{({\rm I})} 
\,,
\end{equation}
where the labels $({\rm S})$ and $({\rm I})$ denote that the quantity they are attached to involves the sensitive body, or only the insensitive bodies, respectively. With $\Phi_2$ split in this way, we now need to generalize the sensitivities that are defined with respect to it. This is achieved by making the following replacement in Equation (\ref{dT}):
\begin{equation}
s_{\Phi_2} \Phi_2 \rightarrow s_{\Phi_2^{({\rm S})}} \Phi_2^{({\rm S})} + s_{\Phi_2^{({\rm I})}} \Phi_2^{({\rm I})} 
\, ,
\end{equation}
which allows our compact body to have different sensitivities to all of the possible contributions to $\Phi_2$. Christoffel symbols and stress-energy components for this geometry, at the required orders, are given in Appendices B and C.

\subsection{Integrability conditions and $\tau^{i0}$ to order $v^5$}

Continuing our calculations to higher order, we now have that the spatial component of Equation (\ref{dt}) evaluated to order $v^6$ can be found to be
\begin{eqnarray} \label{ti6}
\hspace{-2cm}4\pi \, t^{i\mu}_{\phantom{i\mu} ,\mu}  
&=&4 \pi \, a \, U_{,j} t^{ij} + 4 \pi \, \Gamma^i_{\phantom{i} \nu \rho} T^{\nu \rho} +4\pi \, \Gamma^{\nu}_{\phantom{\nu} \nu \rho} T^{\rho i} + 4\pi \, a \, U_{,j} \, \rho \, v^i v^j + 4\pi \, a \, U_{,0} \, \rho \, v^i \\ \nonumber \hspace{-2cm}
&&+ 4 \pi \, a \, U_{,i} \, p - 4\pi s_U \, \rho \, U_{,i} \left(5 \gamma \, U+ \frac{1}{2} v^2 \right) +4\pi \left( s_U^2 +s_U' \right) \rho \, U \, U_{,i} + \sum_i s_{\Phi_i} \Phi_i \, ,
\end{eqnarray}
where the sum at the end of the last line runs over all potentials, including the sensitive and insensitive parts of $\Phi_2$ described above. The first term on the right-hand side of this equation can also be seen to contain $t^{ij}$, but only at order $v^4$, which we already know from Equation (\ref{tij4}). The right-hand side is therefore fully-specified already, and can be manipulated to find $t^{i0}$ at order $v^5$, which is what we will now do.

For this we will use the identities in Appendix D, which can be used to show that
\begin{eqnarray}
\hspace{-2.5cm}4\pi \, \alpha \, \rho^{\star} \, \Phi_{W,j}
&=& - \alpha \, \partial_k \Big[ 2 \Gamma_{jk}(\Phi_W) + 3 U\,  \Gamma_{jk}(U) + 2 \Gamma_{jk} (\nabla X \cdot \nabla U) \\ \hspace{-2.5cm}  &&\nonumber \qquad\qquad - \Sigma_{l,l} \, X_{,jk} - X\, \Sigma_{l,ljk} + \delta_{jk}  (X \, \Sigma_{l,lm})_{,m} +8 \pi \, \rho^{\star} \, X_{,(j}U_{,k)} \Big] \\ \hspace{-2.5cm} \nonumber &&-\partial_j \Big[\left(U_{,0}\right)^2 \Big] + 2 \partial_t \Big[ U_{,j} U_{,0}\Big] +4\pi \, U_{,j} \Big[ \rho^{\star} v^2+2p - \frac{\alpha}{8\pi} \vert \nabla U \vert^2 + \frac{1}{4\pi} \nabla^2 \Phi_6 \Big] \, ,
\end{eqnarray}
where the definitions of $\Gamma_{ij}$ and $\Sigma_i$ are given in the appendix. We now need the right-hand side of Equation (\ref{ti6}) to be written as terms under partial time and space derivatives, and terms which can be written as neither. We note that presenting this equation in this way is {\it not} unique, as there are terms which can be transformed so that they appear under either a time or a space derivative. To remove this ambiguity, we choose to write our expression such that the terms under the time derivative are proportional to either $U_{,i} U_{,0}$ or $U_{,j}V_{[j,i]}$, as these are the form of the terms that appear in $t^{0i}$ in Eq. (\ref{t0i}). We then have the possibility of finding a $t^{i0}$ of the same form by solving Equation (\ref{ti6}), so that we can identify conditions under which $t^{i0}=t^{0i}$, such that globally conserved linear and angular momenta can be discovered.

In this way, the mess of terms after the equality in Equation (\ref{ti6}) can be organized so that they appear in the following way:
\begin{equation} \label{tABC}
4 \pi \, t^{i \mu}_{\phantom{i\mu} ,\mu} = \partial_t A^i + \partial_j B^{ij} + C^i \, ,
\end{equation}
where $C^i$ represents all of the terms that can be written as neither a partial derivative with respect to $t$ nor $x^j$. Equation (\ref{tABC}) is then integrable if $C^i=0$, which in general should be expected to restrict the allowed values of the PPN parameters and our theory-independent sensitivities. In this case, one can identify that $4\pi \, t^{i0} = A^i +O(v^7)$ and $4\pi \, t^{ij}= 4\pi \, t^{(4)ij} +B^{ij} +O(v^8)$, where $t^{(4) ij}$ is the $O(v^4)$ part of $t^{ij}$ from Equation (\ref{tij4}). 

Without making any restrictive assumptions about the vanishing of $C^i$, and using the results from Appendix E, we find that the vector $A^i$ from Equation (\ref{tABC}) can be written as
\begin{eqnarray} \label{Ai}
A^i &=& \Big[ {
\alpha - 5 \gamma +\gamma_1} - 2 \gamma_2 - \frac{2}{\alpha} \beta_W -a - \frac{2}{\alpha} \left( s_{\Phi_5}-s_{\Phi_W} \right) 
\Big] \, U_{,i} \, U_{,0} \\ \nonumber
&&\qquad + \Big[ 2 \alpha - 10 \gamma + 2\gamma_1 - 2 a \Big] \, V_{[j,i]} \, U_{,j} \, ,
\end{eqnarray}
with the skew and symmetric parts of $B^{ij}$ being given by
\begin{eqnarray} \label{Bija}
B^{[ij]} &=& -2 (5\gamma-\alpha+a) \, \partial_t \left(V_{[i,j]} \, U\right)
\end{eqnarray}
and
\begin{eqnarray} 
\hspace{-2cm}B^{(ij)} &=&\Big[ \frac{1}{2} \alpha^2-\beta - \frac{1}{2}\beta_2^{({\rm S})} +\frac{3}{2} \beta_W +\frac{1}{2} a \, \alpha + \alpha \, s_U - \frac{1}{2} a \, s_U \\ \hspace{-2cm}\nonumber &&\hspace{5cm}- \frac{1}{2} s_U^2 -\frac{1}{2} s_U' + \frac{1}{2} s_{\Phi_2^{({\rm S})}} -\frac{3}{2} s_{\Phi_W} \Big] \Gamma_{ij}(U^2) \\ \nonumber
\hspace{-2cm}&&+ 2 \Gamma_{ij} (\psi) + 2 \Big( \beta_W - s_{\Phi_W} \Big) \, \Gamma_{ij} (\nabla X \cdot \nabla U) +2 \gamma_2 \, \Gamma_{ij} ({X}_{,00}) \\ \nonumber
\hspace{-2cm}&&- \left(\beta_W-s_{\Phi_W}\right) \Big[ \Sigma_{k,k} X_{,ij} +X \, \Sigma_{k.kij} - \delta_{ij} (X \, \Sigma_{k,km} )_{,m}\Big] -2 \sum_{k} s_{\Phi_k} \, \Gamma_{ij} (\Phi_k)  \\[-5pt] \nonumber
\hspace{-2cm}&&-4 \gamma_1 \Big[ V_{[i,k]} V_{[j,k]} - \frac{1}{4} \delta_{ij} V_{[m, k]} V_{[m,k]} \Big] - 2 \left( \beta_W - s_{\Phi_W} \right) \, \nabla^2 U \, X_{,(i} U_{,j)}\\ \nonumber
\hspace{-2cm}&&+2 \gamma_1 \Big[ U_{,(i} {V}_{j),0} - \frac{1}{2} \delta_{ij} {V}_{k,0} U_{,k} \Big] - \Big[ \frac{1}{2} \gamma_1 - \gamma_2 -\frac{\beta_W}{\alpha} - \frac{s_{\Phi_5}}{\alpha}  +\frac{s_{\Phi_W}}{\alpha} \Big] \, \delta_{ij} \, {U}_{,0}^2 \\ \nonumber
\hspace{-2cm}&&+ \left( 5 \gamma - \alpha +a \right) \Big[4 \pi \, U\, T_{ij} + \delta_{ij} \, ( U \, {U}_{,0})_{,0} \Big] \, ,
\end{eqnarray}
and with $C^i$ being given by
\begin{eqnarray} \label{Ci}
\hspace{-1.5cm}
C^i &=& \Bigg[-\frac{\alpha}{2} + \beta_1 -\frac{1}{\alpha} {\beta_W}- \gamma -\frac{s_U}{2} - s_{\Phi_1} - \frac{1}{\alpha}s_{\Phi_5} + \frac{1}{\alpha}s_{\Phi_W} \Bigg] \, 4 \pi \, \rho^{\star} \, v^2 \, U_{,i} \\ \hspace{-1.5cm}\nonumber
&&+ \Bigg[ - \beta_6 -\frac{1}{\alpha} {\beta_W} - \frac{1}{\alpha}s_{\Phi_5} +s_{\Phi_6} + \frac{1}{\alpha}s_{\Phi_W} \Bigg] \, U_{,i} \, \nabla^2 \Phi_6 \\ \hspace{-1.5cm}\nonumber &&+ \Bigg[-{\alpha^2} + 2 \beta +{\beta_2^{({\rm S})}} +{\beta_W} -2 \alpha \, s_U +{s_U^2} +{s_U'} -{s_{\Phi_2^{({\rm S})}}} -{s_{\Phi_W}} \Bigg] \, \frac{1}{2} \, U_{,i} \, \vert \nabla U \vert^2 \\ \hspace{-1.5cm}\nonumber
&&+ \Bigg[ -\alpha +\beta_3 - s_{\Phi_3} \Bigg] \, 4\pi \, \rho^{\star} \, \Pi \, U_{,i} \\ \hspace{-1.5cm}\nonumber
&&+ \Bigg[ \beta_4 - \frac{2}{\alpha}{\beta_W} - 3 \gamma - s_{\Phi_4}-  \frac{2}{\alpha}s_{\Phi_5} + \frac{2}{\alpha}s_{\Phi_W}\Bigg] \, 4\pi \, p \, U_{,i} \, .
\end{eqnarray}
From Equation (\ref{tABC}) we can immediately identify {that $4\pi\, t^{i0}= A^i$, which using Equations (\ref{tau}) and (\ref{Ai}) gives $\tau^{i0}$} to order $v^5$ as follows:
\begin{eqnarray} \nonumber
 &&\hspace{-2.5cm}\tau^{i0} = \rho^* v^i \left( 1 {-\tilde{a}U} +\Pi +\frac{1}{2} v^2{+(2 \tilde{\alpha} -3 \tilde{\gamma}) U } \right) +p\, v^i + \frac{1}{4\pi}\big( 2 \tilde{\alpha} - 10 \tilde{\gamma} + 2\gamma_1 - 2 \tilde{a} \big) \, V_{[j,i]} \, U_{,j}
\\ \hspace{-1.5cm} &&+\frac{1}{4\pi}\Big( {\tilde{\alpha} - 5 \tilde{\gamma} +\gamma_1} - 2 \gamma_2 - \frac{2}{\alpha} \beta_W -\tilde{a} - \frac{2}{\alpha} \left( s_{\Phi_5}-s_{\Phi_W} \right) \Big) \, U_{,i} \, U_{,0} \, , \label{taui0}
\end{eqnarray}
while the skew part of $\tau^{ij}$ to order $v^6$ is
\begin{equation} \label{tauija}
\hspace{-1.5cm} \tau^{[ij]} = -\frac{1}{2\pi} (5\tilde{\gamma}-\tilde{\alpha}+\tilde{a}) \, \partial_t \left(V_{[i,j]} \, U\right) \,.
\end{equation}
Equations (\ref{Ci})-(\ref{tauija}), together with Equation (\ref{tau0i}) above, can now be used to identify the conditions required for metric-based theories of gravity to be considered {\it semi-conservative} or {\it fully-conservative} in the presence of a sensitive body.

\vspace{1cm}
\section{Conditions for the existence of global conservation laws} \label{sec:laws}

We will first consider the less restrictive case of semi-conservative theories, before considering the condition for being fully conservative.

\subsection{Semi-conservative theories} \label{sec:semi}

Each of the coefficients inside the square brackets in Equation (\ref{Ci}) must vanish independently, and must do so (in particular) in the absence of sensitivities if the theory is to be at least semi-conservative (i.e. such that $t^{\mu \nu}$ exists at all). To establish the conditions for this we make the following choice for the PPN metric coefficients:
\begin{eqnarray} \label{b1}
\beta_1 &=& \frac{1}{2} ( \alpha +2 \gamma + \alpha_3 + \zeta_1 - 2 \xi)\\
\beta_2^{({\rm I})} &=& (\alpha^2 - 2 \beta + \zeta_2 + \alpha \xi) \\[2pt]
\beta_2^{({\rm S})} &=& (\alpha^2 - 2 \beta + \zeta_2 + \alpha \xi) +f_2 \label{f2}\\[2pt]
\beta_3 &=& \alpha +\zeta_3\\[2pt]
\beta_4 &=& 3 \gamma +3 \zeta_4 - 2 \xi \\
\beta_6 &=& -\frac{1}{2} \zeta_1 + \xi\\
\beta_W &=& -\alpha \xi 
\label{fW} \, .
\end{eqnarray}
These correspond to the usual choices in the PPN approach, with the addition of the extra (as yet unknown) term $f_2$ in the coefficient for $\Phi_2$,  which is included in order to quantify dependence of this potential on the gravitational field of the sensitive body.

Considering first the case in which sensitivities are negligible, we can substitute Equations (\ref{b1})-(\ref{fW}) into our expression for $C^i$ in (\ref{Ci}) to find that each term in that equation vanishes if and only if
\begin{equation} \label{sc0}
\alpha_3 = \zeta_1 = \zeta_2 = \zeta_3 = \zeta_4 = 0 \, , 
\end{equation}
which verifies the usual conditions for a theory of gravity to be semi-conservative. If we now, in addition, require that a theory should remain at least semi-conservative in the presence of sensitivities then we must also require
\begin{eqnarray} \label{sc1}
\hspace{0cm} &&s_{\Phi_3} = 0 \qquad {\rm and} \qquad   s_{\Phi_4} = - 2 s_{\Phi_6} = \frac{2}{\alpha} (s_{\Phi_W} - s_{\Phi_5})= s_U +2  s_{\Phi_1}
\end{eqnarray}
with the additions to the metric coefficient in Equation (\ref{f2}) obeying
\begin{eqnarray} \label{f2r}
f_2 &=& \frac{5}{2} \, \alpha \, s_U - s_U^2 - s_U'  +\alpha \, s_{\Phi_1} +s_{\Phi_2^{({\rm S})}}+s_{\Phi_5}
\end{eqnarray}
This is already a long list of requirements, which leaves only the sets $\{\alpha, \gamma , \xi, {\gamma_1 ,\gamma_2}\}$ and $\{s_U, s'_U, s_{\Phi_1}, s_{\Phi_2^{({\rm S})}}, s_{\Phi_2^{({\rm I})}}, s_{\Phi_5} \}$ as being allowed to be independently specified if a theory is to remain semi-conservative in the presence of a sensitive body.

\subsection{Fully-conservative theories} \label{sec:full}

Assuming the semi-conservative conditions above are met, we also require both $\tau^{0i}=\tau^{i0}$ and $\tau^{[ij]}=0$ in order for a theory to be fully conservative. From Equations (\ref{tau0i}) and (\ref{taui0}), the first of these results in the condition
\begin{equation} \label{fc}
\hspace{-2.5cm}\Big[ {\alpha +2 \gamma -\gamma_1}+2 \gamma_2 +\frac{2}{\alpha} \beta_W + s_U + \frac{2}{\alpha} \left( s_{\Phi_5}-s_{\Phi_W} \right) \Big] \, U_{,i} \, U_{,0}
+ \Big[ 4 \alpha+4 \gamma -2 \gamma_1 \Big] \, V^{[j,i]} \, U_{,j} =0 \,,
\end{equation}
while from Equation (\ref{tauija}) the second results in
\begin{equation} \label{aeq}
(5\gamma-\alpha+a) \, \left( V^{[i,j]} \, U \right)_{,0}= 0 \, .
\end{equation}
On making the usual choices, $\gamma_1 = 2 \alpha+2 \gamma+\frac{1}{2} \alpha_1$ and $\gamma_2 = \frac{1}{2} \alpha +\xi +\frac{1}{2} \alpha_2 - \frac{1}{2} \zeta_1$, these two equations are satisfied in the absence of sensitivities if the PPN parameters obey
\begin{eqnarray}
\alpha_1 &=& 0 \qquad {\rm and} \qquad \alpha_2 = \zeta_1
\end{eqnarray}
and the parameter $a$, introduced in Equation (\ref{tau}), obeys $a = \alpha - 5 \gamma$. With the addition of non-vanishing sensitivity we find that for Equation (\ref{fc}) to be satisfied we must also require
\begin{eqnarray}
\frac{1}{\alpha} \left( s_{\Phi_5}-s_{\Phi_W} \right)   &=& - \frac{1}{2} s_U 
\end{eqnarray}
Taking this condition together with those required to be semi-conservative from Equations (\ref{sc0}) and (\ref{sc1}) means we must require, in total, that
\begin{eqnarray}
\alpha_1 &=& \alpha_2 =\alpha_3 = \zeta_1 = \zeta_2 = \zeta_3 = \zeta_4 = 0 
\end{eqnarray}
and
\begin{eqnarray} \label{fc1}
&&s_{\Phi_1}= s_{\Phi_3}= 0 \qquad {\rm and} \qquad s_{\Phi_4}= -2 s_{\Phi_6}= \frac{2}{\alpha} (s_{\Phi_W} - s_{\Phi_5}) =s_U 
\end{eqnarray}
with
\begin{eqnarray} \label{f2r2}
f_2 &=& \frac{5}{2} \, \alpha \, s_U - s_U^2 - s_U' +s_{\Phi_2^{({\rm S})}} + s_{\Phi_5} 
\end{eqnarray}
Fully conservative theories therefore have only $\{\alpha, \gamma , \xi\}$ and $\{s_U, s_U', s_{\Phi_2^{({\rm S})}}, s_{\Phi_2^{({\rm I})}}, s_{\Phi_5}\}$ as parameters that are allowed to take independent values if a theory is to be fully conservative in the presence of a sensitive body. 

The conditions identified in this section, for theories of gravity to remain either semi or fully-conservative in the presence of sensitivities, appear quite restricting. We will consider special cases in Section \ref{disc}, which will allow for the relaxation of some of them. Before that, we will identify the form of the particular conserved quantities that will exist if these conditions are met in full.

\section{Globally-conserved quantities} \label{sec:quants}

We can now return to the conserved quantities $P^{\mu}$ and $J^{\mu \nu}$. These are obtained by integrating $\tau^{\mu \nu}$ over space, as described in the Equation (\ref{pandj}). For our purposes, it is useful to break this integral up into parts over containing the sensitive and insensitive masses, as follows:
\begin{equation}
P^{\mu}_{({\rm S})} = \int_{\Omega_{\rm S}} \tau^{\mu 0} {\rm d}^3 x \qquad {\rm and} \qquad P^{\mu}_{({\rm I})} = \int_{\Omega_{\rm I}} \tau^{\mu 0} {\rm d}^3 x \, ,
\end{equation}
where $\Omega_{\rm S}$ and $\Omega_{\rm I}$ denote those two regions. Similarly, we can write
\begin{equation}
J^{\mu\nu}_{({\rm S})} = 2 \int_{\Omega_{\rm S}} x^{[\mu} \tau^{\nu]0} \rm{d}^3x
\qquad {\rm and} \qquad
J^{\mu\nu}_{({\rm I})} = 2 \int_{\Omega_{\rm I}} x^{[\mu} \tau^{\nu]0} \rm{d}^3x \, .
\end{equation}
We will study these two types of quantities separately below, inside and outside the sensitive body, before considering them both together.

\subsection{Conserved quantities inside sensitive bodies}

Using our result for $\tau^{00}$ inside the sensitive body (\ref{tau00}), together with the expression for $T^{00}$ from Appendix C, we find that for both semi-conservative and fully-conservative theories we have
\begin{eqnarray}
P^0_{({\rm S})} &=& \int_{\Omega_{\rm S}} \rho  \, \left( 1+ \Pi +\frac{1}{2}v^2 - \frac{1}{2} \, \tilde{\alpha} \, U \right) {\rm d}V + {\rm boundary \; terms}\,,
\end{eqnarray}
where ${\rm d}V= \sqrt{-g}\, u^0 \, {\rm d}^3 x$ is the proper spatial volume element and we note that $\rho$ is now the mass density (i.e. it has no star attached to it). This equation has been derived using integration by parts, and shows that the mass plus internal energy plus kinetic energy plus Newtonian-like gravitational potential energy is a conserved quantity, as long as we calculate the gravitational potential energy using $\tilde{\alpha}= \alpha-s_U$.

For theories that are at least semi-conservative, we have that
\begin{eqnarray}
\hspace{-2cm} P^i_{({\rm S})} &=& \int_{\Omega_{\rm S}} \rho  \, v^i \, \left( 1 + \Pi + \frac{1}{2} v^2 - \frac{1}{2} \, \tilde{\alpha} \, U + \frac{p}{\rho } \right) {\rm d} V  - \frac{1}{2} \int_{\Omega_{\rm S}} \left( \tilde{\alpha}+\alpha_1 \right) \, \rho   \, V^i {\rm d}V\\ \nonumber
\hspace{-2cm}&&\hspace{2.9cm}- \frac{1}{2}\int_{\Omega_{\rm S}} \left( \tilde{\alpha} +{\tilde \alpha}_2 \right) \, \rho  \, X_{,0}^{\phantom{,0}i} {\rm d}V  + {\rm boundary \; terms} \, ,
\end{eqnarray}
where ${\tilde \alpha}_2 \equiv \alpha_2 - 2 s_{\Phi_1}$, and which reduces to the corresponding quantity in the fully-conservative case when $\alpha_1=\alpha_2=s_{\Phi_1}=0$.
In the fully conservative case there also exists the conserved angular and linear momenta:
\begin{eqnarray}
\hspace{-2cm} J^{ij}_{({\rm S})} &=& 2 \int_{\Omega_{\rm S}} x^{[i} v^{j]} \rho  \, \left( 1 + \frac{1}{2} v^2 +\Pi+({\tilde \alpha} + 2 {\tilde \gamma}) \, U + \frac{p}{\rho } \right) {\rm d}V - {\tilde \alpha} \int_{\Omega_{\rm S}} \rho  \, x^{[i} X_{,0}^{\phantom{,0} j]} {\rm d} V \\ \nonumber
\hspace{-2cm}&&\hspace{3.05cm}-4 ({\tilde \alpha} +{\tilde \gamma} ) \int_{\Omega_{\rm S}} \rho  \, x^{[i} V^{j]} {\rm d} V + {\rm boundary \; terms} \\[5pt]
\hspace{-2cm} J^{0i}_{({\rm S})} &=& P_{({\rm S})}^i \, t - \int_{\Omega_{\rm S}} \rho \,  x^i \left( 1+ \frac{1}{2} v^2 +\Pi - \frac{1}{2} {\tilde \alpha} \, U \right) {\rm d}V + {\rm boundary \; terms} \, .
\end{eqnarray}
These are all similar to their usual expressions under the replacements $\{\alpha, \gamma\} \rightarrow \{{\tilde \alpha}, {\tilde \gamma} \}$, and in the semi-conservative case additionally $\alpha_2 \rightarrow {\tilde \alpha}_2$.

\subsection{Conserved quantities outside sensitive bodies}

To calculate these we need to know the gravitational field of sensitive bodies. Using the equivalence of active and passive gravitational masses, it has been argued that this is given in fully-conservative theories by the usual expressions with the replacements  $\alpha \rightarrow {\tilde \alpha}$ and $\gamma \rightarrow {\tilde \gamma}$ \cite{pitt2026constraining} (up to additional higher-order terms in perturbations of the metric). In this case, one should expect the conserved quantities in the region exterior to the sensitive bodies to be given by
\begin{eqnarray}
\hspace{-2cm}P^0_{({\rm I})} &=& \int_{\Omega_{\rm I}} \rho  \, \left( 1+ \Pi +\frac{1}{2}v^2 - \frac{1}{2} \, \tilde{\alpha} \, U^{({\rm S})} - \frac{1}{2} \, {\alpha} \, U^{({\rm I})}\right) {\rm d}V + {\rm boundary \; terms} \\ 
\hspace{-2cm}P^i_{({\rm I})} &=& \int_{\Omega_{\rm I}} \rho  \, v^i \, \left( 1 + \Pi + \frac{1}{2} v^2- \frac{1}{2} \, \tilde{\alpha} \, U^{({\rm S})} - \frac{1}{2} \, {\alpha} \, U^{({\rm I})} + \frac{p}{\rho } \right) {\rm d} V  \\ \nonumber \hspace{-2cm}&&- \frac{1}{2} {\tilde \alpha} \int_{\Omega_{\rm S}}  \rho  \, W^{({\rm S}) i}{\rm d}V - \frac{1}{2} {\alpha} \int_{\Omega_{\rm S}}  \rho  \, W^{({\rm I}) i} {\rm d}V  + {\rm boundary \; terms} 
\end{eqnarray}
and
\begin{eqnarray}
\hspace{-2cm} J^{ij}_{({\rm I})} &=& 2 \int_{\Omega_{\rm I}} x^{[i} v^{j]} \rho  \, \left( 1 + \frac{1}{2} v^2 +\Pi+({\tilde \alpha} + 2 {\tilde \gamma}) \, U^{({\rm S})}+({\alpha} + 2 {\gamma}) \, U^{({\rm I})} + \frac{p}{\rho } \right) {\rm d}V \\ \nonumber
\hspace{-2cm}&&- {\tilde \alpha} \int_{\Omega_{\rm I}} \rho  \, x^{[i} X_{,0}^{({\rm S}) j]} {\rm d} V -4 ({\tilde \alpha} +{\tilde \gamma} ) \int_{\Omega_{\rm I}} \rho  \,x^{[i} V^{({\rm S})j]} {\rm d} V 
\\ \nonumber \hspace{-2cm}&&- {\alpha} \int_{\Omega_{\rm I}} \rho  \, x^{[i} X_{,0}^{({\rm I}) j]} {\rm d} V -4 ({\alpha} +{\gamma} ) \int_{\Omega_{\rm I}} \rho \, x^{[i} V^{({\rm I})j]} {\rm d} V 
+ {\rm boundary \; terms} \\ 
\hspace{-2cm} J^{0i}_{({\rm I})} &=& P_{({\rm I})}^i \, t - \int_{\Omega_{\rm I}} \rho \,  x^i \left( 1+ \frac{1}{2} v^2 +\Pi - \frac{1}{2} {\tilde \alpha} \, U^{({\rm S})}- \frac{1}{2} {\alpha} \, U^{({\rm I})} \right) {\rm d}V + {\rm boundary} \, ,
\end{eqnarray}
where $\{U^{({\rm S})},X^{({\rm S})},V^{({\rm S})i},W^{({\rm S})i} \}$ and $\{U^{({\rm I})},X^{({\rm I})},V^{({\rm I})i},W^{({\rm I})i} \}$ are the potentials associated with the sensitive and insensitive bodies, and where $W^i$ is defined in Appendix A.

\subsection{Conserved global quantities}

We can now construct globally-conserved quantities by linearly combining the contributions from integrals over $\Omega_{\rm S}$ and $\Omega_{\rm I}$, to find
\begin{eqnarray}
\hspace{-2cm}P^0 &=& \int_{\Omega} \rho  \, \left( 1+ \Pi +\frac{1}{2}v^2 - \frac{1}{2} \, \hat{\alpha} \, U \right) {\rm d}V  \\ 
\hspace{-2cm}P^i &=& \int_{\Omega} \rho  \, v^i \, \left( 1 + \Pi + \frac{1}{2} v^2- \frac{1}{2} \, \hat{\alpha} \, U + \frac{p}{\rho } \right) {\rm d} V  - \frac{1}{2} {\hat \alpha} \int_{\Omega} \rho  \, W^{i}{\rm d}V  
\end{eqnarray}
and
\begin{eqnarray}
\hspace{-2cm} J^{ij} &=& 2 \int_{\Omega} x^{[i} v^{j]} \rho  \, \left( 1 + \frac{1}{2} v^2 +\Pi+({\hat \alpha} + 2 {\hat \gamma}) \, U + \frac{p}{\rho } \right) {\rm d}V \\ \nonumber
\hspace{-2cm}&&- {\hat \alpha} \int_{\Omega} \rho  \, x^{[i} X_{,0}^{\phantom{,0} j]} {\rm d} V -4 ({\hat \alpha} +{\hat \gamma} ) \int_{\Omega} \rho  \,x^{[i} V^{j]} {\rm d} V   \\ 
\hspace{-2cm} J^{0i} &=& P^i \, t - \int_{\Omega} \rho \,  x^i \left( 1+ \frac{1}{2} v^2 +\Pi - \frac{1}{2} {\hat \alpha} \, U \right) {\rm d}V  \, ,
\end{eqnarray}
where $\Omega= \Omega_{\rm S}+\Omega_{\rm I}$. The boundary terms between cancel between $\Omega_{\rm S}$ and $\Omega_{\rm I}$ have been taken to cancel, and the boundary terms at spatial infinity have been assumed to vanish. The values of ${\hat \alpha}$ and $\hat{\gamma}$ in these expressions should be taken to correspond to the un-tilde'd quantity when both density and gravitational potential terms in the corresponding part of an integral contain information about insensitive bodies only, and should be taken to be the tilde'd quantity when either corresponds to the sensitive body.

\section{Discussion} \label{disc}

Having derived conditions for globally conserved quantities to exist in the presence of a sensitive body, and having determined the form of the conserved energy, momentum, angular momentum and centre-of-mass momentum, we will now discuss these results. In particular, we will be interested in the simplified scenario where matter is treated as point particles, as well as in comparing our theory-independent approach to a simple example theory, and considering further possible generalizations.

\subsection{Point-particle configurations} \label{sec:pp}

The concept of sensitivities is often applied to situations in which bodies are treated as being simple point-particles. In this case we would normally take each body to have no internal structure, so that internal energy density and pressure vanish: $\Pi=p=0$, which immediately removes the fourth and fifth lines of Equation (\ref{Ci}). The conditions to be semi-conservative (i.e. $C^i=0$) are then satisfied if
\begin{eqnarray} \label{pp}
\frac{1}{\alpha} \left(s_{\Phi_W} - s_{\Phi_5} \right) = -s_{\Phi_6} = \frac{1}{2} s_U + s_{\Phi_1} \, ,
\end{eqnarray}
and if the following conditions on the metric function is met:
\begin{eqnarray}
f_2 &=& \frac{5}{2} \alpha \, s_U - s_U^2 - s_U' +\alpha \, s_{\Phi_1}+s_{\Phi_2^{({\rm S})}}+ \, s_{\Phi_5} \, .
\end{eqnarray}
These seem less restrictive than those of the general case given in Equations (\ref{sc1})-(\ref{f2r}). If, in addition, we also require the theory to be fully-conservative then we also need $s_{\Phi_1}=0$ in the equations above.
The only sensitivities that are prescribed are then $s_{\Phi_1}$ and $s_{\Phi_6}$, and either $s_{\Phi_5}$ or $s_{\Phi_W}$. All other sensitivity parameters are freely specifiable without violating any global conservation laws. This is much less restrictive than the case considered above, but we note that the metric function $f_2$ has to take a very particular form (which we will later see is not always adhered to in simple theories).

We note that the conditions in Equation (\ref{pp}), together with $s_{\Phi_1}=0$ are identical to those obtained by two of us in Reference \cite{pitt2026constraining}, in which two sensitive bodies in circular orbit were modeled as sensitive point particles. In that case these conditions arose from enforcing boost invariance of the modified-EIH Lagrangian, and the potentials $\Phi_5$ and $\Phi_W$ were neglected from the outset, along with their associated sensitivities. From Equation (\ref{pp}) it can be seen that the neglect of the Whitehead potential results in $s_{\Phi_5}=-\frac{1}{2}\alpha\, s_U$ when $s_{\Phi_1}=0$, which means $f_2=2 \alpha \, s_U - s_U^2 - s_U' +s_{\Phi_2^{({\rm S})}}$. Such a requirement (if satisfied) significantly simplifies the modified-EIH Lagrangian, as discussed in Appendix F.

\subsection{A simple example theory: Brans-Dicke gravity} \label{sec:bd}

The simplest alternative theory of gravity is probably the Brans-Dicke theory\footnote{{For another example of an alternative theory that fits into our formalism, see Section 5 of Ref. \cite{forthcoming}.}}, which is a scalar-tensor theory of gravity specified by the action \cite{brans1961mach}
\begin{equation}
S=\frac{1}{16 \pi} \int \left( \varphi R - \frac{\omega}{\varphi} \, \varphi^{,\mu} \varphi_{,\mu} \right) \sqrt{-g} \, {\rm d}^4 x + S_m \, ,
\end{equation}
where $\omega$ is a constant coupling parameter, $S_m$ denotes the action of any matter fields, and $\varphi$ is a new fundamental gravitational field. Extremization of this action results in the field equations
\begin{equation}
R_{\mu \nu} = \frac{8\pi}{\varphi} \left( T_{\mu \nu} - \frac{1}{2} g_{\mu \nu}T \right) +\frac{\omega}{\varphi^2} \phi_{,\mu} \phi_{,\nu} + \frac{1}{\varphi} \left( \varphi_{; \mu\nu} +\frac{1}{2} g_{\mu \nu} \square \varphi \right) \, ,
\end{equation}
with the scalar field propagation equation being given by
\begin{equation} \label{bds0}
\square \varphi = \frac{8\pi \, T}{3+2\omega}  \, .
\end{equation}
We note that these are the full field equations of this theory of gravity, and that the effective theory that one obtains by treating compact bodies as being sensitive to the scalar field $\varphi$ is given by making the replacement $8\pi \, T \rightarrow  8\pi \, T - 16 \pi \, \varphi \, \partial T/ \partial \varphi$ in Equation (\ref{bds0}).

Solving these field equations to post-Newtonian order gives the scalar field as \cite{Mirshekari2013compact, Xie:2007gq}
\begin{equation} \label{bds}
\hspace{-2cm}\varphi = \varphi_0 \left( 1+ 2 \zeta U + 2 \zeta^2 U^2 - 2 \zeta \Phi_2 +4 \zeta^2 s^2 \Phi_2^{({\rm S})}+ 2 \zeta \Phi_3 - 4 \zeta \Phi_4 - \zeta \Phi_5 - \zeta \Phi_6 \right)  \, ,
\end{equation}
where we have neglected terms of order $v^6$ or smaller, and where $\zeta\equiv 1/(4+2 \omega)$. The PPN parameters for this theory are then given by $\alpha=\beta=1$, and $\gamma=1-2 \zeta$ and $\alpha_1=\alpha_2=\alpha_3=\xi=\zeta_1=\zeta_2=\zeta_3=\zeta_4=0$. We can now relate our theory-independent sensitivities to the scalar sensitivities of this theory, which are defined as \cite{eardley1975observable}
\begin{equation}
s\equiv \frac{{\rm d} \ln m}{{\rm d} \ln \varphi} \qquad {\rm and} \qquad s'\equiv \frac{{\rm d}^2 \ln m}{{\rm d} \ln \varphi^2} \, ,
\end{equation}
to find from Equation (\ref{bds}) that
\begin{eqnarray}
\hspace{-1cm} s_U &=& s_{\Phi_3} = -\frac{1}{2} s_{\Phi_4} = -2 s_{\Phi_5} = -2 s_{\Phi_6}=2 \zeta s \, , \qquad s_{\Phi_1} = s_{\Phi_W} = 0 \\ \hspace{-1cm}
s_{\Phi_2^{({\rm I})}} &=& -2 \zeta s \, , \qquad s_{\Phi_2^{({\rm S})}} = -2 \zeta s (1-2 \zeta s) \qquad {\rm and} \qquad s_U' = 4 \zeta^2 s' \nonumber \,,
\end{eqnarray}
as well as the following, which can be obtained from the results of Reference \cite{Mirshekari2013compact}:
\begin{equation}
f_2 = 2 \zeta s\, .
\end{equation}
Substituting all of this into Equation (\ref{Ci}) gives
\begin{eqnarray} 
C_i &=& \zeta s \, \left( 2 \zeta \, \frac{s'}{s} \, \vert \nabla U \vert^2 -8\pi \,   \rho^{\star} \, \Pi  + 24 \pi \,  p \right)  U_{,i} \, .
\end{eqnarray}
This shows that Brans-Dicke theory does not remain conservative if any of $\{\Pi, p,  s' \}$ are non-zero. For a system of point particles, in which $\Pi=p=0$, this reduces to $s'=0$. The fully-conservative requirements that $t^{[0i]}=0 =t^{[ij]}$ are automatically satisfied in these theories, as $s_{\Phi_1}=0$, so if $s'=0$ then $N$-body systems are automatically fully conservative too. We comment further on the relationship between this and the corresponding modified-EIH Lagrangian for these theories in Appendix F.

\subsection{Extended scenarios}

We have so far limited ourselves to considering only a single sensitive body, which is sensitive to only the scalar gravitational potentials of the regular PPN theory. There are various ways that one could extend this scenario, which we will now briefly discuss.

The first generalization, which is of obvious interest, is to extend the analysis presented above so that it includes more than one sensitive body. The case of two sensitive bodies (and no other matter) was considered by two of us in Reference \cite{pitt2026constraining}. In that case the presence of the second sensitive body could be accounted for by generalizing the sensitivity parameters so that they were given to leading order by
\begin{equation}
s_U \rightarrow s^{(1)}_U +s^{(2)}_U - \frac{1}{c_N} s^{(1)}_U s^{(2)}_U = c_N - \frac{1}{c_N} \left( s^{(1)}_U-c_N \right) \left( s^{(2)}_U-c_N \right) \, ,
\end{equation}
where the superscripts identify the body to which the parameter belongs, and where $c_N$ is a constant (the ``critical value'' at which the sensitivity drops out of the gravitational physics). A similar quantity was also constructed at next-to-leading order. To use such expressions in a study of the present form they need to be included in the conservation equations, as in Equation (\ref{dT}), but also in the contributions to the metric from the sensitive bodies; this latter point will become especially important in studies with many bodies, as the metric experienced by any one sensitive body will then depend on the gravitational field produced by the others, as well as all of the insensitive bodies. Including sensitivities in the gravitational fields of compact objects was achieved in Reference \cite{pitt2026constraining} by using symmetry considerations, and by enforcing the equivalence of active and passive gravitational masses. How to achieve this in the presence of more than two sensitive bodies is an open problem, which we leave for the future.

A second generalization of the current study would be to broaden the class of fields to which a compact body can be sensitive. In the work presented above we have allowed the mass of such bodies to be dependent on any of the usual post-Newtonian potentials that are present in the PPN formalism. However, this list is by no means exhaustive (just as in the PPN formalism more generally). It may be that particular theories of gravity require dependence on other fields, or in ways not yet envisaged. We expect that these could be included in generalizations of the present study if/when they are necessary, and will leave this for future work too. For example, beyond the scalar potentials, an obvious addition could be sensitivity due to motion with respect to a preferred frame (in theories where such a thing exists). This has been considered in the context of vector-tensor theories in Reference \cite{taherasghari2025compact}, and combined with scalar sensitivities in Reference \cite{forthcoming}. These studies show that such a sensitivity can act like an environmentally-dependent inertial mass, in addition to the environmentally-dependent gravitational mass that appears in the scalar case.

Other generalizations might include allowing bodies to have internal structure beyond a mass monopole, as well as going to higher orders in the post-Newtonian expansion, and/or including the radiative gravitational sector (to allow the inclusion of gravitational waves). This is left for future work.

\section{Conclusions}

We have investigated the conditions under which global conservation laws exist in theories in which compact bodies are modeled as having a mass density that is sensitive to their local environment. In particular, we have derived explicit conditions that must be obeyed for theories to be semi-conservative, such that they admit globally conserved notions of energy and momentum, as well as being fully-conservative, such that they additionally admit globally conserved angular momentum and centre-of-mass momentum. The conserved quantities themselves have been calculated, and the conditions of existence have been investigated for the special cases of point particles and in a simple example scalar-tensor theory of gravity. Our results are in good keeping with previously known results on this subject, but suggest that some intriguing subtleties may also exist within this subject area.

\section*{Acknowledgements}
We acknowledge support from the Perren Fund, and from STFC under project reference number 2897578.

\section*{Appendix A: Gravitational potentials}

We use the following definitions for gravitational potentials:

\vspace{15pt}
\hspace{1.5cm}
\begin{tabular}{l l }
 $\displaystyle U \equiv \int \frac{\rho^{*\prime}}{\vert {\bf x}-{\bf x'} \vert} \, d^3 x'$ & $\displaystyle X \equiv \int {\rho^{*\prime}}{\vert {\bf x}-{\bf x'} \vert} \, d^3 x'$  \\[15pt]
  $ \displaystyle V^i \equiv \int \frac{\rho^{*\prime} v^{i \prime}}{\vert {\bf x}-{\bf x'} \vert} \, d^3 x'$ & $\displaystyle W^i \equiv \int \frac{\rho^{*\prime} {\bf v^{\prime}} \cdot ({\bf x}-{\bf x'}) (x-x')^i }{\vert {\bf x}-{\bf x'} \vert^3} \, d^3 x'$
\\[15pt] 
  $\displaystyle \Phi_1 \equiv \int \frac{\rho^{*\prime} v^{\prime 2}}{\vert {\bf x}-{\bf x'} \vert} \, d^3 x'$ & 
 $\displaystyle \Phi_2 \equiv \int \frac{\rho^{*\prime} U^{\prime}}{\vert {\bf x}-{\bf x'} \vert} \, d^3 x'$ 
 \\[15pt] $\displaystyle \Phi_3 \equiv \int \frac{\rho^{*\prime} \Pi^{\prime}}{\vert {\bf x}-{\bf x'} \vert} \, d^3 x'$ &
 $\displaystyle \Phi_4 \equiv \int \frac{p^{\prime}}{\vert {\bf x}-{\bf x'} \vert} \, d^3 x'$ 
 \\[15pt]  $\displaystyle \Phi_5 \equiv \int \rho^{*\prime} \nabla'U' \cdot \frac{({\bf x}-{\bf x'})}{\vert {\bf x}-{\bf x'} \vert} \, d^3 x'$ &
 $\displaystyle \Phi_6 \equiv \int \rho^{*\prime} \, \frac{\left[{\bf v'} \cdot({\bf x}-{\bf x'})\right]^2}{\vert {\bf x}-{\bf x'} \vert^3} \, d^3 x'$  \\[15pt]
 $\displaystyle \Phi_W \equiv \int \hspace{-5pt}\int \rho^{*\prime} \rho^{*\prime\prime} \, \frac{({\bf x}-{\bf x'})}{\vert {\bf x}-{\bf x'} \vert^3} \cdot \left[
\frac{({\bf x'}-{\bf x''})}{\vert {\bf x}-{\bf x''} \vert}
- \frac{({\bf x}-{\bf x''})}{\vert {\bf x'}-{\bf x''} \vert}
\right] \, d^3 x' d^3 x''$ \hspace{-6cm}\phantom{a} & 
\end{tabular}
\vspace{15pt}

\noindent
where a prime on a function denotes its dependence on ${\bf x'}$, such that e.g. $\rho^{*\prime}\equiv\rho^*(t,{\bf x'})$, and where $\Pi$ and $p$ are the internal energy density and pressure of the matter, respectively. The index on $V^i$ can be lowered and raised with the Kronecker delta and its inverse, as is usual in relativistic perturbation theory.

\section*{Appendix B: Christoffel symbols}

The Christoffel symbols derived from the metric with components (\ref{g00})-(\ref{gij}) are given, to the required order, by

\begin{eqnarray}
\hspace{-1.5cm}
\Gamma^0_{\phantom{0} 00} &=& - \alpha U_{,0} +O(v^4) \, , \qquad \Gamma^j_{\phantom{j} 0k} = \gamma \delta_{jk} U_{,0} - \gamma_1 V_{[j,k]} +O(v^4) \\ \hspace{-1.5cm}
\Gamma^0_{\phantom{0} 0j} &=& -\alpha U_{,j} +O(v^3) \, , \qquad \Gamma^j_{\phantom{j} kl} = \gamma \left( \delta_{jk} U_{,l} +\delta_{jl} U_{,k} - \delta_{kl} U_{,j} \right) +O(v^3)  \\ \hspace{-1.5cm}
\Gamma^0_{\phantom{0} jk} &=& \gamma \delta_{jk} U_{,0} +\gamma_1 V_{(j,k)} + \gamma_2 X_{,0jk} +O(v^4) \\ \hspace{-1.5cm}
\Gamma^j_{\phantom{j} 00} &=& -\alpha U_{,j} +2 (\beta+ \alpha \gamma) U U_{,j} - \psi_{,j} - \gamma_1 V_{,0j} - \gamma_2 X_{,j00} +O(v^5)
\,.
\end{eqnarray}

\section*{Appendix C: Stress-energy components}
The components of the stress-energy tensor are, to the required orders, given by
\begin{eqnarray}
T^{00} &=& \rho^{\star} \left[ 1 +\Pi + \frac{1}{2} v^2 +(2 \alpha - 3 \gamma+s_U) U \right] +O(v^6) \\
T^{0j} &=& \rho^{\star} v^j \left[ 1+\Pi +\frac{1}{2}v^2+(2\alpha-3\gamma+s_U) U \right]+ p\, v^j + O(v^7) \\[5pt]
T^{jk} &=& \rho^{\star} v^j v^k + p \,\delta^{jk} +O(v^6) \, .
\end{eqnarray}

\section*{Appendix D: Useful identities}

The following identities are useful for performing the calculations in this paper:
\begin{eqnarray}
U_{,j} \Gamma_{ij} (U) &=& \frac{1}{2} U_{,i} \vert \nabla U \vert^2 \\
4\pi \, \rho^{\star} \, U U_{,i} &=& -\frac{1}{2} \partial_j \Gamma_{ij}(U^2) + \frac{1}{2} U_{,i} \vert \nabla U \vert^2 \\
\Gamma_{ij}(U^2) &=& 2 U \,\Gamma_{ij}(U) \\
4 \pi \, \rho^{\star} \, \Phi_{1,i} &=& - 2 \partial_j \Gamma_{ij}(\Phi_1) - 4\pi \, \rho^{\star} \, v^2 \, U_{,i} \\
4 \pi \, \rho^{\star} \, \Phi_{2,i} &=& - 2 \partial_j \Gamma_{ij}(\Phi_2) + \frac{1}{2} \partial_j \Gamma_{ij} (U^2) - \frac{1}{2} U_{,i} \vert \nabla U \vert^2\\
4 \pi \, \rho^{\star} \, \Phi_{3,i} &=& - 2 \partial_j \Gamma_{ij}(\Phi_3) - 4 \pi \, \rho^{\star} \, \Pi \, U_{,i}\\
4 \pi \, \rho^{\star} \, \Phi_{4,i} &=& - 2 \partial_j \Gamma_{ij}(\Phi_4) - 4 \pi \, p \, U_{,i}\\
X_{,00} &=& \Phi_1+2 \Phi_4 - \alpha \, \Phi_5 - \Phi_6\\
\nabla^2 \Phi_W &=& -4 \vert \nabla U \vert^2 + 4 \pi \, \rho^{\star} \, U - 2 U_{,ij} X_{,ij}\\
\Sigma_{i,i} &=& \frac{1}{2} \nabla^2 \Phi_5
\end{eqnarray}
and
\begin{eqnarray}
\rho^{\star} \, v^i \, \frac{{\rm d} U}{{\rm d} t} &=& \partial_t \left( \rho^{\star} \, v^i \, U\right) + \partial_j \left( \rho^{\star} \, v^i \, v^j \, U \right) - \tilde{\alpha} \, \rho^{\star} \, U \, U_{,i} +  p_{,i} \, U\\
\rho^{\star} \, \frac{{\rm d} V^i}{{\rm d} t} &=& \partial_t \left( \rho^{\star} \, V^i \right) + \partial_j \left( \rho^{\star} \, v^j \, V^i \right)
\end{eqnarray}
where
\begin{eqnarray}
\frac{{\rm d} }{{\rm d} t} &\equiv& \frac{\partial}{\partial t} + v^i \, \frac{\partial}{\partial x^i}\\
\Gamma_{ij}(f) &\equiv& U_{,(i}f_{,j)} - \frac{1}{2} \delta_{ij} \nabla U \cdot \nabla f\\
\nabla^2 \Sigma_{i} &\equiv& -4\pi \, \rho^{\star} \, U_{,i} \, .
\end{eqnarray}

\newpage

\section*{Appendix E: Sensitive and insensitive parts of $\Phi_2$}

The parts of $\Phi_2$ that correspond to the sensitive body appearing in its definition, or insensitive bodies only, obey the following within a sensitive body:
\begin{eqnarray}
4\pi \, \rho^{\star} \, \Phi_{2,i}^{({\rm S})} &=& -2 \partial_j \Gamma_{ij} \left(\Phi_2^{({\rm S})} \right) - 4 \pi \, \rho^{\star} \, U \, U_{,i} \\
4\pi \, \rho^{\star} \, \Phi_{2,i}^{({\rm I})} &=& -2 \partial_j \Gamma_{ij} \left( \Phi_2^{({\rm I})} \right)
\end{eqnarray}
where
$$
\Phi_{2}^{({\rm S})} = \int \hspace{-5pt}\int \frac{\rho^{\star \prime}}{\vert {\bf x}-{\bf x'}\vert} \frac{\rho^{\star \prime \prime}_{({\rm S})}}{\vert {\bf x'}-{\bf x''} \vert} {\rm d}^3 x'{\rm d}^3 x'' \quad {\rm and} \quad \Phi_{2}^{({\rm I})} = \int \hspace{-5pt}\int \frac{\rho^{\star \prime}}{\vert {\bf x}-{\bf x'}\vert} \frac{\rho^{\star \prime \prime}_{({\rm I})}}{\vert {\bf x'}-{\bf x''} \vert} {\rm d}^3 x'{\rm d}^3 x''
$$
and where $\rho^{\star}_{({\rm S})}$ and $\rho^{\star}_{({\rm I})}$ correspond to the conserved mass densities of sensitive and insensitive bodies, respectively.

\section*{Appendix F: Comparison to modified EIH parameters}

The Einstein, Infeld and Hoffmann (EIH) Lagrangian is a result from general relativity that one can use to calculate the motion of massive bodies \cite{einstein1938gravitational}. This approach has been generalized by Will to include other theories of gravity by the construction of a modified-EIH Lagrangian \cite{will2018testing}:
\begin{eqnarray} \label{eih}
    L_{{\rm EIH}} &=& - \sum_{a} m_{a} \left[ 1 - \frac{1}{2} v_{a}^{2} - \frac{1}{8} (1 + \mathcal{A}_{a}) v_{a}^{4} \right] \\ \nonumber
    &&+ \frac{1}{2} \sum_{a} \sum_{b \neq a} \frac{m_{a} m_{b}}{r_{ab}} \left[ \mathcal{G}_{ab} + 3 \mathcal{B}_{ab} v_{a}^{2} - \frac{1}{2}(\mathcal{G}_{ab} + 6 \mathcal{B}_{(ab)} + \mathcal{C}_{ab}) \mathbf{v}_{a} \cdot \mathbf{v}_{b} \right. \\ \nonumber
    &&\left. - \frac{1}{2} (\mathcal{G}_{ab} + \mathcal{E}_{ab}) (\mathbf{v}_{a} \cdot \mathbf{n}_{ab}) (\mathbf{v}_{b} \cdot \mathbf{n}_{ab}) \right]  - \frac{1}{2} \sum_{a} \sum_{b \neq a} \sum_{c \neq a} \mathcal{D}_{abc} \frac{m_{a} m_{b} m_{c}}{r_{ab} r_{ac}} \, ,
\end{eqnarray}
where the parameters $\{\mathcal{A}_a , \mathcal{B}_{ab} , \mathcal{C}_{ab} , \mathcal{D}_{abc} , \mathcal{E}_{ab} , \mathcal{G}_{ab} \}$ are to depend on the parameters of theory under consideration. Each body $a$ depends on a mass parameter $m_a$ and a 3-velocity $\mathbf{v}_a$, with $r_{ab}$ being the distance between bodies $a$ and $b$ and $\mathbf{n}_{ab}$ being a unit vector from $a$ to $b$. The parameters $\{\mathcal{C}_{ab} , \mathcal{E}_{ab} , \mathcal{G}_{ab} , \mathcal{D}_{cab}\}$ are automatically symmetric under the interchange $a \leftrightarrow b$, with $\mathcal{B}_{ab}$ being symmetric and $\{\mathcal{A}_{a} ,\mathcal{C}_{ab} , \mathcal{E}_{ab} \}$ all being required to vanish if the theory has no preferred frame.

The values of the modified-EIH parameters have been calculated in two-body systems in the presence of theory-independent sensitivities by two of us in Reference \cite{pitt2026constraining}, under the assumptions that the gravitational fields of the bodies pressure, $p$, and internal energy, $\Pi$, are negligible (as described in Section \ref{sec:pp}) and that the bodies are only sensitive to the potentials $ \{U, \Phi_1, \Phi_2, \Phi_6 \}$. Under this specific set of assumptions, the results found in Section \ref{sec:semi} that are required for a theory to be semi-conservative in the presence of sensitivities, as specified in Equations (\ref{sc1}) and (\ref{f2r}), can be been seen to be equivalent to requiring $\mathcal{B}_{12}=\mathcal{B}_{21}$ and $\mathcal{D}_{122}=\mathcal{D}_{211}$ respectively, where $1$ and $2$ label the two bodies in the system. This appears to show that the semi-conservative requirement is satisfied if and only if there is a symmetry between the sensitive and insensitive bodies in the kinematics of the system, suggesting that the semi-conservative conditions require the gravitational field of the sensitive body to influence the motion of the insensitive body in an exactly equivalent form to the way that the motion of the sensitive body responds to the gravitational field of its insensitive counterpart. This seems to be a new and interesting correspondence between the concepts of semi-conservatism and reciprocity of active and passive gravitational masses, in the presence of sensitivities. Going further, the conditions to be fully conservative from Section \ref{sec:full}, as given in Equations (\ref{fc1})-(\ref{f2r2}), are then equivalent to the condition to have no preferred frames that was found in Reference \cite{pitt2026constraining}. That is, both of these conditions require the additional constraint $s_{\Phi_1}=0$ to be satisfied. This verifies that the relationship between fully-conservative theories and frame-independent theories is preserved in the presence of sensitive matter, at least under the conditions of these studies.

With regard to the scalar-tensor theory considered in Section \ref{sec:bd}, we note that a modified-EIH Lagrangian can still be constructed even when $s'\neq 0$ \cite{will2018testing}, which nevertheless implies conserved quantities despite the failure of these theories to be even semi-conservative in this case. This highlights a key difference between the possible existence of conserved quantities in the equations of motion of $N$-body systems, and the existence of integral conservation laws of the types derived in Section \ref{sec:quants}: the former is possible without the latter. Indeed, the conditions we have derived to be semi-conservative have already been seen to impose additional requirements on our theory-independent sensitivities, beyond those required for a modified-EIH Lagrangian to exist, and these scalar-tensor theories show a direct realization of this idea. That is, when $s'=0$ we have the semi-conservative conditions being satisfied {\it and} the result that $\mathcal{D}_{122} = \mathcal{D}_{211}$ in a two-body system, and when $s' \neq 0$ we have neither of these things. In this latter case the theory is not semi-conservative and two-body modified-EIH Lagrangian is not invariant under $1 \leftrightarrow 2$. This happens despite the existence of conserved quantities in the equations of motion of such systems, which come about from the existence of a generating Lagrangian. In this case, such quantities will be constructed by combining the effects of the gravitational field $U^2$ of the sensitive body with the response of the sensitive body to the gravitational potential $\Phi_2$ of the insensitive body. Such a combination cannot occur when constructing integral conserved quantities, of the type found in Section \ref{sec:quants}, as they require combining effects at different spatial locations rather than simply integrating a quantity that obeys a local conservation equation (such as $\tau^{\mu \nu}$ in Equation \ref{pandj}).

\newpage

\section*{References}
\bibliographystyle{ieeetr}


\end{document}